# Optical forces, helicity, angular momentum and how they are all intertwined

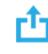

View Online    Export Citation    CrossMark

Iker Gómez-Viloria,[1,a] 🔟 Enrique Ayllón García,[1] 🔟 Jorge Olmos-Trigo,[1] 🔟 Quimey Pears Stefano,[1] 🔟
Jon Lasa-Alonso,[1,2,3] 🔟 Martín Molezuelas-Ferreras,[1] 🔟 and Gabriel Molina-Terriza[1,2,4,b] 🔟

## AFFILIATIONS

[1] Centro de Física de Materiales (CFM), CSIC-UPV/EHU, Paseo Manuel de Lardizabal 5, 20018 Donostia-San Sebastián, Spain
[2] Donostia International Physics Center, Paseo Manuel de Lardizabal 4, 20018 Donostia-San Sebastián, Spain
[3] Basic Sciences Department, Mondragon Unibertsitatea, Olagorta Kalea 26, 48014 Bilbao, Spain
[4] IKERBASQUE, Basque Foundation for Science, Maria Diaz de Haro 3, 48013 Bilbao, Spain

**Note:** This paper is part of the Special Topic on Angular Momentum of Light.
[a] Author to whom correspondence should be addressed: iker_gomez@hotmail.com
[b] Electronic mail: gabriel.molina.terriza@gmail.com

## ABSTRACT

The theoretical description of optical forces and torques on micron-sized particles is a crucial area of research and has formed the foundation for advancements in optical trapping and manipulation technologies. In this study, we derive analytical expressions for optical forces and torques on micron-sized spherical particles illuminated by focused Laguerre–Gaussian (LG) beams, employing the well-defined helicity multipolar decomposition of electromagnetic fields and Mie theory. We developed a multifunctional program, Multipolar Optical Forces Toolbox, based on this theoretical framework. The program, available on GitHub, was used to generate optical trapping stability maps. These maps predict trap stability across a wide range of system parameters and serve as a practical tool for designing advanced optical trapping experiments. Our analysis reveals the important role of helicity $p$ and orbital angular momentum $\ell$ on the dynamics of particles trapped off-axis in LG beams and demonstrates the unique nature of the tangential torque. Our findings also highlight notable differences in longitudinal optical forces resulting from pure helicity modifications in Gaussian beams. Furthermore, we showcase the ability of LG beams to isolate Mie resonances, offering a novel approach to locate the spectral positions of the resonances of high multipolar modes. These insights deepen the understanding of helicity in LG optical traps and pave the way for the development of more advanced optical manipulation techniques.

## I. INTRODUCTION

The study of optical forces and torques exerted on micrometer-sized particles by light fields has been a subject of intense research in recent decades.[1] These forces play a crucial role in various fields such as optical trapping, manipulation, and sorting of particles, as well as in fundamental studies of light–matter interactions. Understanding and accurately characterizing these forces is essential for the development of advanced optical manipulation techniques and their applications in diverse areas ranging from biophysics to nanotechnology.[2,3]

Optical trapping, in particular, has emerged as a powerful tool for particle manipulation, offering precise control over particles ranging from nanometers to micrometers in size. Since the first experiments demonstrating the feasibility of trapping particles using light,[4–6] optical tweezers have revolutionized various research fields, including biology,[2] biochemistry,[3] and even enabling the cooling of atoms to ultra-low temperatures.[7,8]

Initial optical trapping techniques were based on the use of tightly focused Gaussian beams. However, several advances in optical manipulation have extended beyond Gaussian beams to more complex structured light fields, such as Laguerre–Gaussian (LG) beams. These beams, characterized in the paraxial regime by their helical wavefronts with topological charge number $\ell = \ldots, -2, -1, 0, 1, 2, \ldots$, carry a customizable value of orbital angular momentum, offering unique opportunities to control the optical

28 January 2026 19:28:14

forces and torques exerted on particles. Moreover, while focusing LG beams with circular polarization, their cylindrical symmetry is preserved, allowing alternative and insightful analyses of the optical trapping system.[9] In this scenario, the beams can be described with the helicity $p = \pm 1$ and the total angular momentum in the direction of propagation ($z$ axis) $m_z = \ell + p$.[10,11] These quantities are useful in the description of beams under paraxial propagation, in the highly focused regime and, also, when scattering of cylindrically symmetric samples.[12] These features enable unique control over optical forces and torques, giving rise to advancements in particle manipulation techniques.

In the mid-1990s, researchers began exploring the transfer of angular momentum from light beams with helical wavefronts to particles. Professor Rubinsztein-Dunlop and co-workers demonstrated the transfer of orbital angular momentum from LG beams to small particles.[13] They subsequently showed the transfer of spin angular momentum using circularly polarized beams.[14] These experiments were further explored by Padgett and colleagues, who developed setups to transfer both spin and orbital angular momenta to birefringent particles, effectively creating an "optical wrench" or "optical spanner."[15] In their experiment, a particle trapped in the bright ring of an LG beam was rotated either around the beam's center in an orbital motion or around the particle's axis in a spin motion, by adjusting the $\ell$ and $p$ parameters of the trapping beam.

Understanding the behavior of particles illuminated by LG beams requires a theoretical framework capable of accurately capturing the intricate light–matter interactions. While traditional optical force calculations through ray-optics models[16–19] offer insights by discretizing the incident light beam into a bundle of rays, this approach is only valid for very large particles and incident beams with no phase singularities. To overcome this drawback, Generalized Lorenz–Mie Theory (GLMT) emerges as a reliable method for calculating optical forces, providing exact solutions to the scattering problem of spherical particles under general illumination conditions.[20]

Electromagnetic fields can be represented with the multipolar decomposition, which uses an orthonormal basis of spherical multipolar vectors that are solutions to the Maxwell equations. In this framework, each multipole is weighted by coefficients known as Beam Shape Coefficients (BSCs). For paraxial beams, various methods can be used to calculate the BSC agreeing remarkably well with several experimental trapping beams.[21–23] However, for non-paraxial beams, only a few solutions for BSC can be found in the literature, and they have not yet been fully exploited in modeling optical trapping systems, likely due to their recent formulation.[24,25] All these methods are based on numerical calculations that often make the physical intuition of the phenomena at hand quite elusive. However, the correct application of these novel methods to represent realistic non-paraxial beams is crucial for an accurate modeling of optical trapping systems, especially since highly focused beams are the ones predominantly used in optical trapping experiments. In this regard, the use of high numerical aperture objectives is essential for creating high light intensity gradients, enabling stable 3D optical trapping of particles. In this paper, we show that the development of a theoretical model for optical forces based on beams of well-defined helicity not only provides a better fit for comparison with experimental results but also offers a good intuition on how optical forces, helicity, and angular momentum are intertwined.

In our theoretical framework, we describe the incident and scattered electromagnetic fields using the well-defined helicity multipolar decomposition, formed as a linear combination of electric and magnetic multipoles.[11,26] In addition, the resulting BSCs are used to reach one of the most realistic mathematical descriptions of the electromagnetic fields present in optical trapping systems using focused LG-modes. We start deriving the BSC for a beam in an "on-focus" configuration, where LG-modes impinge on the center of spherical particles. Then, these coefficients are modified in order to describe displaced "off-focus" LG beams in terms of the well-defined helicity multipoles, allowing precise descriptions of different optical trapping situations. Applying Mie's theory, we obtain exact solutions for the scattering problem of spherical particles under general illumination conditions. To compute the optical forces and torques, we utilize this multipolar description of light fields and the integration of the Maxwell Stress Tensor (MST),[27,28] offering an exact analytical solution of the optical forces in terms of the BSC. The accuracy of our theoretical framework compared to other approximate methods is similar to the one provided by the most realistic descriptions of non-paraxial beams of GLMT,[29–31] which is also based on multipolar expansion of electromagnetic fields, but using the well-known electric and magnetic multipolar modes. In addition, the reliability of the specific procedure described in this work has already been successfully demonstrated in Ref. 9, where the experimental and numerical measurements of the trap stiffness constant were compared, showing an excellent agreement. With all this, our theoretical framework offers an accurate computation of the optical forces, which is particularly useful for highly focused circularly polarized beams. Moreover, it provides a deeper understanding of the transfer of spin and orbital angular momenta to the trapped particles.

In this paper, we provide analytical expressions for the optical forces and torques, considering the well-defined helicity multipolar decomposition of the incident-focused light field and the Mie coefficients defined by the intrinsic properties of the spherical particle. This model allows us to accurately analyze the optical forces and torques exerted on spherical particles by LG beams, enabling novel studies regarding the role of the angular momentum that they carry. Based on this theoretical model, we built the GitHub project Multipolar Optical Forces Toolbox (MOFT), which simulates LG optical traps and reveals particle dynamics, optical forces, and torques. Thanks to this tool, the trapping stability maps shown in this work were created, providing a comprehensive analysis of the optical trap stability across a wide range of system parameters. These stability maps offer valuable predictions and guidance for designing future optical trapping experiments. Our study reveals the intricate interplay between the helicity and orbital angular momentum of the incident beam in shaping the dynamics of particles optically trapped in LG beams, particularly in off-axis configurations. Through our analysis of the symmetries of this kind of optical systems, we uncover the exceptional nature of the tangential torque, revealing its dependence on the topological charge $\ell$ and helicity $p$. We demonstrate significant differences in longitudinal optical forces arising uniquely from flipping the helicity of Gaussian beams. Furthermore, we explore the capability of LG beams to isolate Mie resonances through the observation of longitudinal optical forces, providing an alternative method to determine the spectral positions of higher-order multipolar resonances. Overall, our findings offer valuable insights into the role of helicity in LG optical traps and pave the way for the development of advanced optical manipulation techniques.

## II. ELECTROMAGNETIC FIELDS IN A WELL-DEFINED HELICITY BASIS

For our definitions of the electromagnetic fields, we follow the ones given by Rose in Eqs. (2.63) and (2.64) of Ref. 32. As defined in these references, any incident electromagnetic field can be described as a superposition of electric ($e$) and magnetic ($m$) multipoles, i.e., $\mathbf{A}_{j,m_z}^{(m/e)(n)}$, where the index $n = 1, 3$ refers to the radial function of the multipolar modes being spherical Bessel functions. The remaining indices, $j$ and $m_z$, are associated with the angular momentum carried by these multipoles: first, they are eigenmodes of the squared total angular momentum operator $\mathbf{J}^2$, being $j(j + 1)$ the corresponding eigenvalue, and, second, they are eigenmodes of the $z$ component of the total angular momentum operator $J_z$, with $m_z$ being the corresponding eigenvalue. The amplitudes of the multipolar modes of the incident field are typically called Beam Shape Coefficient (BSC) and are denoted by $g_{j,m_z}^{(e)}$ and $g_{j,m_z}^{(m)}$[28] (see Appendix A for more details),

$$\mathbf{E}_{in} = E_0 \sum_{j=1}^{\infty} \sum_{m_z=-j}^{j} g_{j,m_z}^{(m)} \mathbf{A}_{j,m_z}^{(m)(1)} - g_{j,m_z}^{(e)} \mathbf{A}_{j,m_z}^{(e)(1)}. \tag{1}$$

From this expression and using Mie theory, one obtains the field scattered by a spherical object as

$$\mathbf{E}_{sc} = -E_0 \sum_{j=1}^{\infty} \sum_{m_z=-j}^{j} g_{j,m_z}^{(m)} b_j \mathbf{A}_{j,m_z}^{(m)(3)} - g_{j,m_z}^{(e)} a_j \mathbf{A}_{j,m_z}^{(e)(3)}, \tag{2}$$

where $a_j$ and $b_j$ are the Mie coefficients, whose expressions can be found in Appendix A.

However, here, we will focus on the description of the incident and scattered fields in terms of multipolar fields with well-defined helicity:[26,32,33] $\mathbf{A}_{j,m_z}^{p\,(n)}$, which are the eigenstate of the helicity operator for monochromatic fields $\Lambda = (1/k)\nabla\times$,[11] with $p$ being the corresponding eigenvalue of helicity. In terms of electric and magnetic modes, the helicity multipolar modes read as

$$\mathbf{A}_{j,m_z}^{p\,(n)} = \frac{\mathbf{A}_{j,m_z}^{(m)(n)} + ip\mathbf{A}_{j,m_z}^{(e)(n)}}{\sqrt{2}}. \tag{3}$$

Helicity is defined as the projection of the angular momentum onto the direction of the linear momentum, and, in the case of paraxial beams, it matches directly with their circular polarization state. When a paraxial beam is focused, it acquires different propagation directions and we can no longer define a polarization state. Under typical experimental conditions, circularly polarized beams that are tightly focused can be approximated as having a well-defined helicity value, since most microscope objective lenses are designed to have the same transmissivity of ordinary and extraordinary polarizations of the incoming paraxial light. This can also be seen as the preservation of the helicity of the beam during focusing.[12] Therefore, helicity is an interesting quantity for describing such beams and strengthen the suitability of our theoretical framework. In addition, the $\mathbf{A}_{j,m_z}^{p\,(n)}$ modes offer certain advantages in the analytical development of the scattering theory and the optical forces, as we will show below.

We can then derive the expressions of the incident electric field carrying well-defined helicity and the corresponding scattered field using the multipoles $\mathbf{A}_{j,m_z}^{p\,(n)}$, by defining new BSCs $g_{j,m_z,p}$,

$$\mathbf{E}_{in} = E_0 \sum_{j=1}^{\infty} \sum_{m_z=-j}^{j} g_{j,m_z,p} \mathbf{A}_{j,m_z}^{p\,(1)}, \tag{4}$$

$$\mathbf{E}_{sc} = E_0 \sum_{j=1}^{\infty} \sum_{m_z=-j}^{j} g_{j,m_z,p} \left( \alpha_j \mathbf{A}_{j,m_z}^{p\,(3)} + \beta_j \mathbf{A}_{j,m_z}^{-p\,(3)} \right), \tag{5}$$

where

$$\alpha_j = -\frac{a_j + b_j}{2}, \qquad \beta_j = \frac{a_j - b_j}{2}. \tag{6}$$

Here, the electric field amplitude $E_0$ can be written in terms of the optical power of the incident beam $P_{beam}$ as

$$E_0 = k\sqrt{8ZP_{beam}}, \tag{7}$$

where $Z = \sqrt{\mu/\varepsilon}$ is the impedance of the electromagnetic wave, with $\varepsilon$ being the permittivity and $\mu$ being the permeability of the medium.

It is important to note that, in general, even if the incident field contains only one helicity component, the scattered electric field would, in general, contain both negative and positive helicity contributions. This occurs because a sample does not generally preserve the incident helicity after scattering. Helicity conservation occurs only if the sample is dual.[34] By definition, a dual sample exhibits identical electric and magnetic responses, thereby preserving the polarization state of the incoming beam. Consequently, when a dual object is illuminated by a beam with well-defined helicity, helicity remains as a conserved quantity after scattering. For the scattering of a homogeneous dielectric sphere, electromagnetic duality is reached if $a_j = b_j \forall j$.[34] However, it has been proved that for non-magnetic homogeneous spheres, achieving duality is, in general, impossible.[35] In this paper, we will mostly focus on incident fields of well-defined helicity, i.e., either $p = 1$ or $p = -1$.

## III. BEAM SHAPE COEFFICIENTS

There are two main advantages of using helicity multipolar modes for calculating scattering fields and optical forces: (1) The expressions of the BSCs are very simple in most of the usual experimental configurations, and (2) therefore, they provide a physical intuition of the phenomena involved.

Let us dwell first on the role of the helicity in scattering or optical trapping experiments. Often, one would start with a paraxial beam, which can be modified with standard techniques (polarization and spatial control, etc.). This beam is subsequently focused with a microscope objective in order to achieve higher spatial resolution, increased signal, or a tighter optical trap. Furthermore, in optical trapping, one needs to determine the scattering for displacements of the scatterer in the field of the trap. The interesting fact is that helicity is conserved both in an ideal microscope objective (for example using the typical aplanatic approximation)[12] and in displacements.[36,37] This means that if an experimentalist starts with a paraxial beam with circular polarization, i.e., $\mathbf{E}_{par} = E(x, y)\boldsymbol{\xi}_p$, with $\boldsymbol{\xi}_p = -p(\hat{\mathbf{x}} + ip\hat{\mathbf{y}})$, as defined in Appendix A, then the focused and displaced beams would only contain multipoles with helicity $p$ with the same value as the previous polarization state. Notice that a focused beam would, in general, have polarization components $\boldsymbol{\xi}_0 = \hat{\mathbf{z}}$ and $\boldsymbol{\xi}_{-p}$ but still will have just a single helicity component.[11]

As any paraxial field can be expressed as a superposition of circular polarizations, it suffices to form an intuition of what happens with fields of well-defined helicity and then observe further effects due to interference when using arbitrarily polarized incident paraxial fields. Even so, in this work, we will mainly focus on beams with well-defined helicity.

### A. On-focus beam shape coefficients for LG beams

We start by deriving the Beam Shape Coefficients (BSCs) for an on-focus Laguerre–Gaussian (LG) beam carrying well-defined helicity and focused by an aplanatic lens. First, we assume that the LG beam is paraxial before reaching the lens, which is an appropriate approach in many typical experimental scenarios. Importantly, paraxial circularly polarized LG beams have a well-defined value of total angular momentum in their direction of propagation ($z$ axis), which in this work is referred to as $m_z^*$ and can be calculated as the sum of the orbital and spin angular momenta of the LG beam in that direction. The point is that, in this kind of beams, these quantities are directly defined by their topological charge $\ell = \ldots, -2, -1, 0, 1, 2, \ldots$ and helicity $p = \pm 1$, correspondingly. Here, as mentioned earlier, the helicity $p$ has the same value as the polarization state of the beam. Thereby, the total angular momentum in the direction of propagation carried by a paraxial circularly polarized LG beam is $m_z^* = \ell + p$.

Now, let us analyze these properties if the LG beam is focused by a lens. Assuming a perfect alignment between the beam axis and the center of the lens, the focused beam would preserve the cylindrical symmetry around $z$, and, as a consequence, the total angular momentum in $z$ of the paraxial beam would be conserved in the focusing process.[10,11] Moreover, as discussed earlier, we can also assume that the LG beam preserves its helicity when it is focused by the lens, which is the typical case for microscope objectives.[12] Then, the helicity $p$ in the focused part of the LG beam would be the same value as the circular polarization state of the paraxial part. Under these usual focusing assumptions, we can state that $m_z^*$ and $p$ are well-defined values in focused LG beams whose paraxial part was circularly polarized.

The mathematical description of these non-paraxial LG beams can be expressed in terms of well-defined helicity multipoles, considering that the beam is focused at the origin of coordinate system (on-focus configuration). As reasoned earlier, these kinds of beams only contain multipolar modes with fixed values $m_z^* = \ell + p$ and $p = \pm 1$ and are described as

$$\mathbf{E}_{in}^{on} = E_0 \sum_{j=|m_z^*|}^{\infty} g_{j,m_z^*,p}^{on} \mathbf{A}_{j,m_z^*}^p. \tag{8}$$

Here, the superscript "on" has been added to differentiate the decomposition of an on-focus beam from the one of a displaced off-focus beam, which will be done in Sec. III B. Moreover, given that $m_z^*$ represents the value of the component of the angular momentum in the direction $z$, the total angular momentum, expressed in terms of $j$, cannot have a smaller value. This follows from the well-known relation for the angular momentum $j \geq |m_z^*|$[38,39] and determines the lower limit of the summation over $j$ in Eq. (8), setting it to $|m_z^*|$. The calculation of $g_{j,m_z^*,p}^{on}$ is done by using the Aplanatic Lens Model (ALM),[40–42] as it is described in Refs. 11, 24, 26, and 33. For simplicity, we separate the BSCs into two parts: $g_{j,m_z^*,p}^{on} = D_j C_{j,m_z^*,p}^{on}$, where the weights $D_j = i^j (2j + 1)^{\frac{1}{2}}$ are the standard multipolar coefficients of plane waves. Now, $C_{j,m_z^*,p}^{on}$ for a focused paraxial field $\mathbf{E}_{par} = E_0 f_\ell(\rho) \exp(i\ell\phi)\boldsymbol{\xi}_p$ ($\rho$ and $\phi$ are cylindrical coordinates in the transversal plane before the microscope objective) is

$$C_{j,m_z^*,p}^{on} = \int_0^{\theta_k^M} \sin\theta_k \, d\theta_k \, f \, e^{-ikf} \sqrt{\pi \cos(\theta_k)} \, d_{m_z^*,p}^j(\theta_k) f_\ell(f \sin\theta_k), \tag{9}$$

where the maximum half-angle of collection from the lens is $\theta_k^M = \arcsin(\mathrm{NA}/n_m)$, with NA being its nominal numerical aperture and $n_m$ being the refractive index of the medium after the focusing lens [hence, it coincides with the one that is used in the Mie coefficients of Eq. (A6)]. $f$ is also related to the nominal NA and aperture diameter $D_l$ as $f = n_2 D_l/(2\mathrm{NA})$. The $d_{m_z^*,p}^j(\theta_k)$ are the Wigner small delta. If $E(\rho\cos(\phi), \rho\sin(\phi)) = f_\ell(\rho)\exp(i\ell\phi)$ is normalized to 1 in the transversal plane, then $\sum \|g_{j,m_z^*,p}^{on}\|^2 = 1$, and the electric field amplitude $E_0$ can be written in terms of the optical power of the incident beam $P_{beam}$ in Eq. (7). In our case, we will mainly use a set of Laguerre–Gaussian beams, i.e., $f_\ell(\rho) = LG_{\ell,q}(\rho)$. The definition and normalization of paraxial Laguerre–Gaussian modes can be found in Appendix B.

### B. Off-focus beam shape coefficients

As mentioned earlier, in some situations, such as in optical trapping, it is interesting to calculate the scattering of a sphere displaced with respect to the focal point of the beam, a configuration also known as "off-focus." Again, the multipoles of well-defined helicity offer a convenient way of calculating these coefficients. In the case of spherical objects, it is more convenient to maintain the scatterer on the center of coordinates. In this way, the Mie coefficients of the beam remain unaltered. Meanwhile, the BSCs of the beam will be modified: $g_{j,m_z,p}^{off,\mathbf{d}} = D_j C_{j,m_z,p}^{off,\mathbf{d}}$. Considering an arbitrary displacement of the beam, the beam will, in general, lose its cylindrical symmetry, and its BSCs may contain components of $m_z$ other than the initial $m_z^*$. However, helicity commutes with translations,[37] which means that, if the initial on-focus beam has a well-defined helicity $p$, the displaced beam will have the same helicity components. Then, the displaced beam will be

$$\mathbf{E}_{in}^{off}(\mathbf{d}) = E_0 \sum_{j=1}^{\infty} \sum_{m_z=-j}^{j} g_{j,m_z,p}^{off,\mathbf{d}} \mathbf{A}_{j,m_z}^p, \tag{10}$$

where $\mathbf{d} = d\hat{\mathbf{d}}$ represents the vectorial displacement of the beam focus. The analytical expression of the $C_{j,m_z,p}^{off,\mathbf{d}}$ coefficient has been derived in Refs. 36 and 37 and has the following form:

$$C_{j,m_z,p}^{off,\mathbf{d}} = \sum_{j'=|m_z^*|}^{\infty} \mathscr{T}_{j,m_z}^{j',m_z^*}(k\mathbf{d}) \, C_{j',m_z^*,p}^{on},$$

$$\mathscr{T}_{j,m_z}^{j',m_z^*}(k\mathbf{d}) = (-1)^{m_z^*-m_z} \sum_{L=0}^{\infty} (2L+1)(-i)^L j_L(kd) \mathscr{D}^L(\hat{\mathbf{d}})_0^{m_z^*-m_z} \\ \times \langle j, m_z; L, m_z^* - m_z | j', m_z \rangle \langle j, p; L, 0 | j', p \rangle,$$ (11)

where $\mathscr{T}_{j,m_z}^{j',m_z^*}(k\mathbf{d})$ is the displacement matrix, which depends on $j_L(kd)$, the spherical Bessel function; $\langle j_1, m_1; j_2, m_2 | J, M \rangle$, a Clebsch–Gordan coefficient; and $\mathscr{D}^J(\hat{\mathbf{r}})_m^n$, the Wigner D-matrix. An intuitive picture of how these translation matrices operate arises from recognizing that, in the helicity basis, electromagnetic fields are composed of circularly polarized plane waves (both propagating and evanescent). Thus, translating the field does not affect the polarization state of such plane waves with respect to their direction of propagation. Meanwhile, Clebsch–Gordan coefficients emerge from the coupling of states with different angular momentum. In other words, given two states $|j_1, m_1\rangle$ and $|j_2, m_2\rangle$, Clebsch–Gordan coefficients make the resulting coupled state an eigenstate of the total angular momentum operators. Moreover, Wigner D-matrices can be intuitively understood as the complex amplitudes needed to express a plane wave in terms of spherical harmonics, and vice versa. In particular, D-matrices with $n = 0$ are proportional to the spherical harmonics $Y_{jm}(\hat{\mathbf{r}})$.[37] Notice that the translation matrices can be computed beforehand and stored in order to speed up the numerical calculations. In addition, note that translations fulfill the group operation $\hat{\mathscr{T}}(\mathbf{x}_1 + \mathbf{x}_2) = \hat{\mathscr{T}}(\mathbf{x}_1)\hat{\mathscr{T}}(\mathbf{x}_2)$, where matrix multiplication is understood, meaning that given a certain precision of the calculation, only a handful of displacement matrices are needed. It is important to note that the normalization of $C_{j,m_z,p}^{off,\mathbf{d}}$ is the same as the one of $C_{j,m_z^*,p}^{on}$, so the amplitude of the electric field $E_0$ can be calculated using Eq. (7) as well.

Finally, notice that for the specific case, when a displacement is performed only along the $z$ axis, the system remains cylindrically symmetric with respect to this axis with the consequences mentioned in Sec. III A. In these cases, $\mathscr{T}_{j,m_z}^{j',m_z^*}(kd\hat{\mathbf{z}}) \propto \delta_{m_z^*}^{m_z}$, and we will refer to such optical systems as "on-axis," considering that the focal point of the beam is positioned along the $z$ axis. Consequently, displaced LG beams with a focal point located away from the $z$ axis will be referred to as "off-axis" throughout this Tutorial.

## IV. OPTICAL FORCES IN TERMS OF THE WELL-DEFINED HELICITY COEFFICIENTS

Here, we provide the analytical solution to the optical forces in terms of the well-defined helicity coefficients. We consider an arbitrary incident field and the scattered field, first from an arbitrary scatterer and then particularizing for a spherical object. The analytical derivation of the time-averaged optical force $\langle \mathbf{F} \rangle$ comes from the Minkowski form of the Maxwell Stress Tensor (MST) $\overleftrightarrow{\mathbb{T}}$, integrating it over a spherical shell surrounding the illuminated particle.[27] This operation represents the conservation of the electromagnetic wave momentum principle and is expressed as

$$\langle \mathbf{F} \rangle = \oint_S \overleftrightarrow{\mathbb{T}} \cdot \hat{\mathbf{n}} \, d\mathbb{S},$$ (12)

where $\hat{\mathbf{n}}$ is a unitary vector that is normal to the differential of the area of a sphere surrounding the sample $d\mathbb{S}$ and

$$\overleftrightarrow{\mathbb{T}} = \frac{1}{2} \text{Re}\left[ \varepsilon \, \mathbf{E}_{tot} \otimes \mathbf{E}_{tot}^* + \mu \, \mathbf{H}_{tot} \otimes \mathbf{H}_{tot}^* - \frac{1}{2}\left( \varepsilon |\mathbf{E}_{tot}|^2 + \mu |\mathbf{H}_{tot}|^2 \right) \overleftrightarrow{\mathbb{I}} \right]$$ (13)

is the Minkowski form of the MST,[43] where the total electromagnetic fields are considered, calculated as $\mathbf{E}_{tot} = \mathbf{E}_{in} + \mathbf{E}_{sc}$ and $\mathbf{H}_{tot} = \mathbf{H}_{in} + \mathbf{H}_{sc}$.

Since the integral over the sphere surrounding the scatterer can be chosen arbitrarily, one can only consider the far-field contribution of the total electromagnetic fields. In this case, the optical forces simplify, resulting in

$$\mathbf{F} = -\frac{r^2}{4} \text{Re} \int \left( \varepsilon |\mathbf{E}_{tot}|^2 + \mu |\mathbf{H}_{tot}|^2 \right) \hat{\mathbf{n}} \, d\Omega,$$ (14)

where $d\Omega = \sin\theta \, d\theta \, d\varphi$ denotes the differential solid angle and $\hat{\mathbf{n}}$ is a unitary vector that is normal to the differential of the area of a sphere surrounding the sample $d\mathbb{S}$. Note that the time average of the force is assumed.

Now, we expand the electromagnetic fields in multipolar modes with well-defined helicity. Taking into account the most generic case in which the incident field carries both helicities and separating the total electric field in positive ($p = +1$) and negative ($p = -1$) helicity multipoles, we have that

$$\mathbf{E}_{tot} = E_0 \sum_{j,m_z,p} \left[ A_{j,m_z,p} \, \mathbf{A}_{j,m_z}^{p\,(1)} + B_{j,m_z,p} \, \mathbf{A}_{j,m_z}^{p\,(3)} \right],$$ (15)

where

$$A_{j,m_z,p} = g_{j,m_z,p} = D_j C_{j,m_z,p}$$ (16)

and $B_{j,m_z,p}$ are the amplitudes of the components of the scattered field. For the particular case of a spherical scatterer, they take a very simple expression,

$$B_{j,m_z,p} = \alpha_j \, A_{j,m_z,p} + \beta_j \, A_{j,m_z,-p} = D_j \left( \alpha_j \, C_{j,m_z,p} + \beta_j \, C_{j,m_z,-p} \right).$$ (17)

The integrals resulting from the expressions of the helicity multipoles are analytical,[27,28] and the resulting expressions are

$$F_z = -\frac{|E_0|^2 \varepsilon}{2 k^2} \sum_{j,m_z,p} \sqrt{w_j} \frac{w_{j,m_z}^{(1)}}{j+1} \, \text{Im}\Big[ A_{j+1,m_z,p} B_{j,m_z,p}^* \\ - A_{j,m_z,p} B_{j+1,m_z,p}^* + 2 \, B_{j+1,m_z,p} B_{j,m_z,p}^* \Big] \\ + \frac{p \, m_z}{j(j+1)} \Big[ \text{Re}\Big( A_{j,m_z,p} B_{j,m_z,p}^* \Big) + |B_{j,m_z,p}|^2 \Big]$$ (18)

and

$$
\begin{bmatrix} F_x \\ F_y \end{bmatrix} = \begin{bmatrix} \mathrm{Re} \\ \mathrm{Im} \end{bmatrix} \frac{i|E_0|^2 \varepsilon}{4 \, k^2} \sum_{j,m_z,p} \sqrt{w_j} \, \frac{w^{(2)}_{j,m_z}}{j+1} \left( A_{j+1,m_z,p} B^*_{j,m_z+1,p} \right.
$$
$$
+ \, B_{j+1,m_z,p} A^*_{j,m_z+1,p} + 2 \, B_{j+1,m_z,p} B^*_{j,m_z+1,p} \Big)
$$
$$
+ \, \sqrt{w_j} \, \frac{w^{(3)}_{j,m_z}}{j+1} \left( A_{j,m_z,p} B^*_{j+1,m_z+1,p} \right.
$$
$$
+ \, B_{j,m_z,p} A^*_{j+1,m_z+1,p} + 2 \, B_{j,m_z,p} B^*_{j+1,m_z+1,p} \Big)
$$
$$
+ \, p \, \frac{i \, w^{(4)}_{j,m_z}}{j(j+1)} \left( A_{j,m_z,p} B^*_{j,m_z+1,p} + B_{j,m_z,p} A^*_{j,m_z+1,p} \right.
$$
$$
\left. + 2 \, B_{j,m_z,p} B^*_{j,m_z+1,p} \right), \tag{19}
$$

where $w_j = j(j+2)/(2j+3)(2j+1)$, $w^{(1)}_{j,m_z} = \sqrt{(j+m_z+1)(j-m_z+1)}$, $w^{(2)}_{j,m_z} = \sqrt{(j-m_z+1)(j-m_z)}$, $w^{(3)}_{j,m_z} = \sqrt{(j+m_z+2)(j+m_z+1)}$, and $w^{(4)}_{j,m_z} = \sqrt{(j-m_z)(j+m_z+1)}$.

A few remarks are in order so that we can provide some intuitive understanding of these expressions. One can observe how the expression for the forces naturally separates into two helicity components that do not mix,[44] i.e., there is no term such as $A_p B_{-p}$ or $B_p B_{-p}$. This can be compared with the use of electric and magnetic multipolar modes,[27,28,45,46] where the optical forces mix the electric and magnetic components. However, one can observe interference effects in the forces when using an incident field with a superposition of both helicities if the scatterer is not dual, i.e., if $\beta_j \ne 0$ in the case of spherical scatterers. In this case, Eq. (17) shows that the amplitudes and phases of the helicity components can have different effects on the forces. Finally, as can be expected from the usual calculations of optical forces, the only terms that matter in the optical forces are the scattering terms of the $B_p B_p$ kind and the interference of the scattering and incident fields ($A_p B_p$). (See Appendix C to find the expressions of the longitudinal and transversal optical forces for spherical particles and incident beams carrying well-defined helicity.)

## V. OPTICAL TORQUES IN TERMS OF THE WELL-DEFINED HELICITY COEFFICIENTS

Equivalently to the optical force calculation, the conservation principle of the electromagnetic wave angular momentum can be used in order to obtain the spin optical torques exerted on a particle. The tensor that represents the electromagnetic angular momentum flux at every point of space is $\overleftrightarrow{\mathbb{M}} = \mathbf{r} \times \overleftrightarrow{\mathbb{T}}$, with $\mathbf{r}$ being the position vector where the angular momentum flux of the electromagnetic field is evaluated.[47] Hence, the spin torque exerted on the illuminated particle would be

$$
\langle \mathbf{T} \rangle = \oint_S \overleftrightarrow{\mathbb{M}} \, d\mathbb{S}. \tag{20}
$$

The calculations of the torque using multipolar fields of well-defined helicity can be most easily carried out following Ref. 45. Here, we give the expressions of the optical torques for spherical objects, resulting in

$$
T_z = -\frac{|E_0|^2 \varepsilon}{4 \, k^3} \sum_{j,m_z,p} m_z |A_{j,m_z,p}|^2 \left[ |\alpha_j|^2 + |\beta_j|^2 + \mathrm{Re}(\alpha_j) \right], \tag{21}
$$

for the $z$ component, and also for the $x$ and $y$ directions,

$$
\begin{bmatrix} T_x \\ T_y \end{bmatrix} = -\begin{bmatrix} \mathrm{Re} \\ \mathrm{Im} \end{bmatrix} \frac{|E_0|^2 \varepsilon}{4 \, k^3} \sum_{j,m_z,p} w^{(4)}_{j,m_z} A_{j,m_z,p}
$$
$$
\times A^*_{j,m_z+1,p} \left[ |\alpha_j|^2 + |\beta_j|^2 + \mathrm{Re}(\alpha_j) \right]. \tag{22}
$$

Notice that, as expected, due to the rotational symmetry of the spheres, all the expressions depend only on $\left[ |\alpha_j|^2 + |\beta_j|^2 + \mathrm{Re}(\alpha_j) \right]$, which is proportional to the absorption cross section of the particle, which is only non-zero if the sphere absorbs light. The optical torque is nonzero if and only if the sphere is lossy, and this holds regardless of the beam's position relative to the sphere. This can be formulated as follows: Even if the system loses cylindrical symmetry, a sphere will not rotate around its axis unless it is lossy.

## VI. INSIGHTS INTO THE ROLE OF THE HELICITY IN LAGUERRE–GAUSSIAN OPTICAL TRAPS

The correct and efficient characterization of the dynamics of optically trapped particles in Laguerre–Gaussian modes, such as in the experiments of Ref. 9, has posed a challenging task for researchers in this field. In this section, we present various results obtained through our optical force and torque calculation method, demonstrating its significant potential. Furthermore, the well-defined helicity basis employed in this method allows for a straightforward analysis of the relationships between optical forces and torques and the polarization state of the incoming beam. The results shown in this section have been calculated using the Python program of the GitHub repository "MOFT—Multipolar Optical Forces Toolbox"[48] within the "QNanoLab" organization.

### A. Stability maps of optical traps with LG modes

As an example of the versatility of our open-source code, let us start by studying the stability of optical trapping with focused Laguerre–Gaussian (LG) modes in an on-axis configuration. Our code allows us to calculate efficiently the stability of optical traps, measured as the minimum depth of the potential wells generated by optical forces, for various common optical trapping scenarios. For the study performed in this section, in specific cases where the trapping regime results in an off-axis configuration, the stability of the trap is considered nonexistent or zero. This is because off-axis optical traps using LG beams do not provide a stationary trapping regime, causing the orbital rotation of the particle around the beam axis.

First, we examine the case of trapping a particle with low refractive index contrast, specifically $SiO_2$ in water (index contrast of $m \sim 1.1$). Figure 1 illustrates the stability of the on-axis optical trap cases produced by focused LG modes with topological charges of $\ell = 0$ and $\ell = 1$, acting on a spherical $SiO_2$ particle at a wavelength of $1 \, \mu m$. The stability is represented in panels (a) and (b) using a color map that shows different numerical apertures (NAs) of the focusing lens and varying optical sizes $x = k \, a$ of the particle, with $a$ being its radius. In all calculations, the effects of buoyant and gravitational

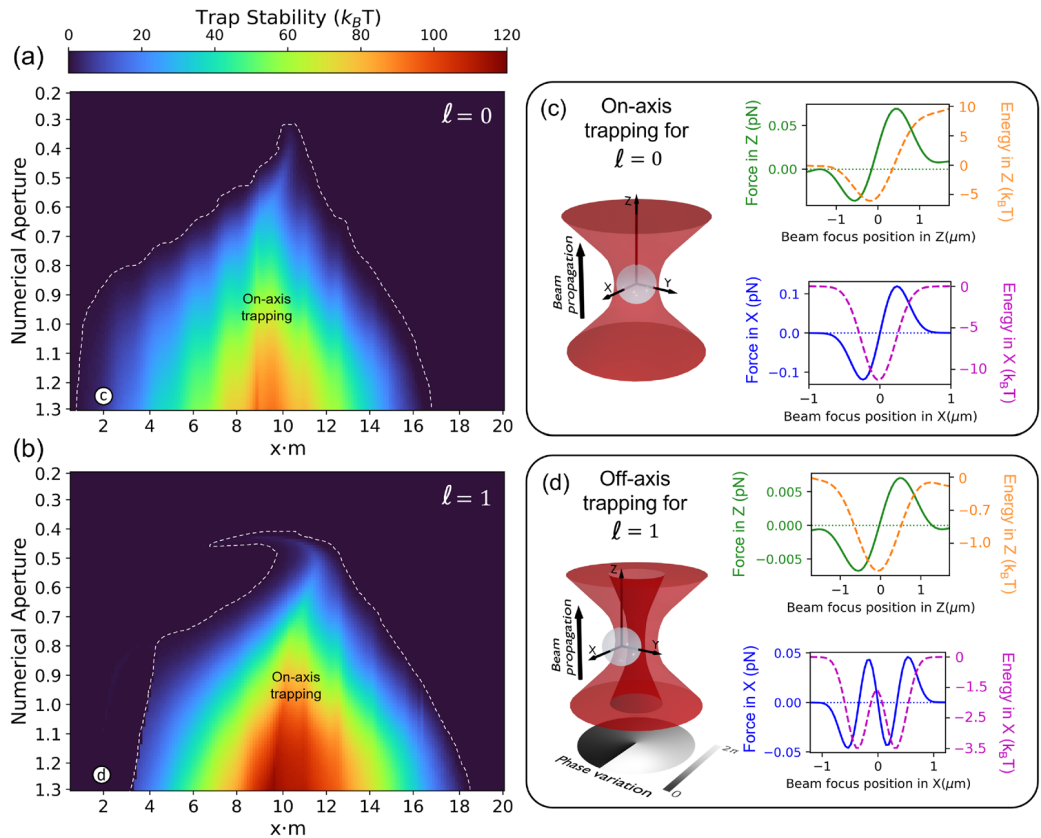

**FIG. 1.** (a) and (b) Trapping stability of a SiO₂ particle in water illuminated by a Laguerre–Gaussian beam with helicity $p = 1$ and topological charge $\ell = 0$ (a) and $\ell = 1$ (b), for varying numerical apertures and optical sizes ($x \cdot m$). The white dashed line indicates the limits of on-axis trapping. (c) and (d) Force calculation for the two topological charges, $\ell = 0$ (c) and $\ell = 1$ (d), at the position indicated in (a) and (b), respectively. At the left of each panel, there is a schematic representation of the focused Laguerre–Gaussian beam and the silica particle trapped on-axis for $\ell = 0$ (c) and off-axis for $\ell = 1$ (d) At the right, the longitudinal total force (green solid line) and its trap energy (orange dashed line) as a function of the displacement in Z, and the transverse optical force (blue solid line) and its trap energy (purple dashed line) as a function of the displacement in X. Here, the transition from an on-axis trapping (c) to the off-axis trapping (d) mentioned in the text can be seen. The power of the beam before going through the objective lenses is 1 mW, and the wavelength is 1 $\mu$m.

forces were included in the computation of the longitudinal forces, considering a vertical trap.

Both LG modes of Fig. 1 exhibit a wide range of conditions favorable for optical trapping. When comparing the results of (a) and (b), the first thing one observes is the higher trapping stability of the $\ell = 1$ mode, as already reported.[9,17,49–51] Another key distinction is that the trapping range (marked with a white line) for $\ell = 1$ extends to larger optical sizes ($x \cdot m > 17$). However, for smaller sizes ($x \cdot m < 3$), on-axis trapping is lost, transitioning to off-axis trapping,[52–54] where the particle is trapped in the bright ring of the LG beam. This feature is highlighted in panels (c) and (d). In each panel, a diagram on the left illustrates the corresponding Laguerre–Gaussian focused beam along with the trapped silica particle, while on the right, the longitudinal (along the $z$ axis) and transverse (along the $x$ axis) total forces are shown as solid lines, accompanied by their corresponding trapping energies, represented by dashed lines, for the parameters $x \cdot m = 2$ and NA = 1.25. The $z$ axis forces (green line) demonstrate stable trapping for both topological charges, but the $x$ axis forces reveal that, for the $\ell = 1$ mode, the trapping is not centered, indicating a loss of stable on-axis trapping. The on-axis and off-axis trapping regimes can be easily distinguished by observing the position of the energy minima along the $x$ axis (dashed purple): on-axis trapping occurs when the energy minimum is centered at $x = 0$ Fig. 1(c), whereas off-axis trapping occurs when the minima are located at $x \neq 0$ [Fig. 1(d)]. In Secs. VI B and VI C, we will further explore the differences between on-axis and off-axis trapping in terms of spin and orbit torques.

We also studied the case of optical trapping for high refractive index contrast particles, focusing on the stability of TiO₂ in vacuum (index contrast of $m \sim 2.5$). Figure 2 shows the stability of an on-axis optical trap produced by the same LG modes than before, but for a TiO₂ spherical particle at a wavelength of 1 $\mu$m. The stability maps are also depicted using a color map in panels (a) and (b) across different numerical apertures and optical sizes. For these calculations, we considered a vertical trap, so the gravitational force was included

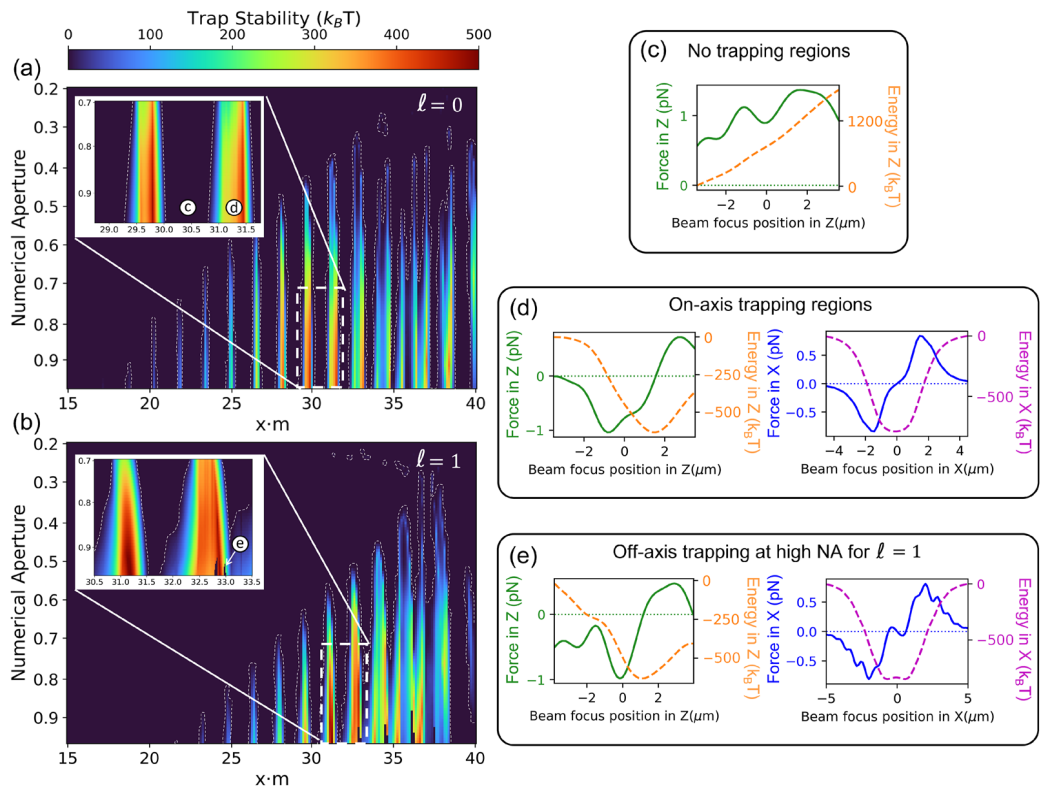

FIG. 2. (a) and (b) Trapping stability of a TiO$_2$ particle in air, illuminated by a Laguerre–Gaussian beam with helicity $p = 1$ and topological charge $\ell = 0$ (a) and $\ell = 1$ (b), for varying numerical apertures and optical sizes ($x \cdot m$). The white line indicates the limits of on-axis trapping (c) Longitudinal force in the periodic no-trapping region marked in the zoom of (a). (d) Longitudinal force (left) and transverse (right) forces in the periodic trapping region marked in the zoom of (a). (e) Longitudinal force (left) and transverse (right) forces in the region marked in the zoom of (b), where off-axis trapping occurs. The power of the beam before passing through the objective lenses is 1 mW, and the wavelength is 1 $\mu$m.

in the computation of the longitudinal forces and trap stability maps.

In this case, the trapping regions (marked with a white line) are considerably fewer compared to silica, but an interesting fringe pattern is observed. For both LG modes, trapping cannot be achieved for small optical sizes ($x \cdot m < 20$). However, for larger optical sizes, narrow, periodic trapping zones appear, extending to low numerical apertures. Two of these trapping regions are shown zoomed in the insets of the figures, showing a width of approximately $\Delta x \cdot m \simeq 0.5$. This peculiar behavior arises from the oscillation of the longitudinal optical force with the optical size, which will be analyzed in Sec. VI E. When varying the optical size, two scenarios arise: one in which the optical force is at a maximum, and thus no trapping is achieved (c), and another in which the optical force reaches a minimum and is countered by the particle's weight, creating a stable trapping scenario (d). Finally, for the $\ell = 1$ mode, we observe that at high NA, in certain small regions, on-axis trapping is lost and transitions to off-axis trapping, making it unstable (e).

These simulations demonstrate the potential of the developed theory and our freely available code[48] to make theoretical predictions for various optical trapping scenarios. Moreover, the results obtained from the simulations of TiO$_2$ in vacuum offer significant potential for controlling the trapping and release of particles by merely adjusting the wavelength of the incident beam by a few nanometers. In addition, this approach may enable selective trapping of particles within a narrow size distribution, while repelling those with optical sizes outside this range. This will be further explored in Sec. VI D in the context of Mie resonances.

## B. The singular nature of tangential torque in off-axis LG optical traps

Here and in Sec. VI C, we will further explore the differences between on-axis and of-axis trapping with Laguerre–Gaussian (LG) modes. Indeed, in this subsection, the nature of off-axis optical traps with LG beams is further explored. In Fig. 3, we show the influence of the sign of the helicity $p$ and topological charge $\ell$ of an incident LG beam on the dynamics of an optically trapped spherical particle. In the presented results, we consider an absorbing particle with a diameter of 0.5 $\mu$m and a complex refractive index of $1.5 + 0.001i$, resulting in an absorption coefficient $\alpha_{abs} = 236.20$ cm$^{-1}$. The rest of the parameters of the optical systems shown in Fig. 3 are common, and they have been meticulously selected in order to generate off-axis optical trapping simultaneously in each of the four cases. That is, the particle is trapped along the brightest part of the annular beam and experiences orbital rotation around the beam axis. This movement is generated by a constant force acting tangentially to the circumference of the beam near the focal plane, with a rotation direction defined by the sign of the paraxial beam topological charge $\ell$. Figure 3 shows the instantaneous trapping position when the beam focus is displaced in the $x/z$ plane, making the optical forces in these two axes zero ($F_x = F_z = 0$). Hence, at that beam position, the tangential force acting on the particle is solely in the $y$ direction, as observed in Fig. 3. Notice that the magnitude of forces depends on the absolute value of $m_z^* = \ell + p$, which is a relevant quantity in focused beams. Indeed, we observe a different force for a fixed $\ell$ and different $p$, which changes the total angular momentum from $|m_z^*| = 3$ to $|m_z^*| = 1$. From the experimental point of view, flipping the circular polarization of the incident paraxial field will generally have an effect on the optical forces for a given azimuthal phase $\ell$.

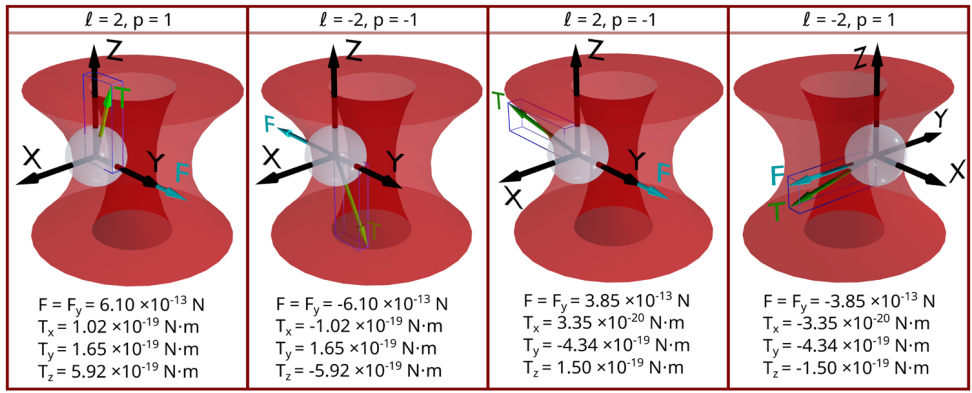

**FIG. 3.** Qualitative representation of optical forces and torques acting on a spherical particle undergoing off-axis optical trapping in highly focused Laguerre–Gaussian beams. The simulations consider a spherical particle of 0.5 $\mu$m diameter and refractive index $n = 1.5 + 0.001i$ surrounded by water and Laguerre–Gaussian laser beams with wavelength $\lambda = 532$ nm and 30 mW of optical power, focused with NA = 1.3. This figure illustrates the four distinct trapping scenarios resulting from the possible sign combinations of the topological charge $\ell \pm 2$ and helicity $p \pm 1$ of the incident beam.

Furthermore, in Fig. 3, we recognize that the absolute value of the optical torques remains constant for $\ell = \pm 2$, $p = \pm 1$, as well as for $\ell = \pm 2$, $p = \mp 1$, due to the symmetries of the optical systems presented. This is because the torque on the sphere only depends on the total angular momentum of the modes centered at the position of the sphere. Interestingly, the directions of the torques exhibit peculiar orientations: if we analyze the signs of the different torque and force components in each case, we have

$$\ell = \pm 2, p = \pm 1 \rightarrow \pm F_y, \pm T_x, +T_y, \pm T_z$$
$$\text{and} \quad\quad\quad\quad\quad\quad\quad\quad\quad\quad\quad (23)$$
$$\ell = \pm 2, p = \mp 1 \rightarrow \pm F_y, \pm T_x, -T_y, \mp T_z.$$

It is noticeable that the orientation of the tangential component of the torque, which in this situation is $T_y$, is affected by the sign modification of $\ell$ and $p$ in a different way than the rest of the components. In particular, $T_y$ is the only component that remains constant while performing a simultaneous flip of the sign for $\ell$ and $p$.

This is due to the symmetry transformations of angular momentum with respect to rotations and parity. More concretely, we can understand the behavior of $T_y$ by considering the transformation that changes a beam with $\ell = 2$, $p = 1$ to $\ell = -2$, $p = -1$, and similarly, a beam with $\ell = 2$, $p = -1$ to $\ell = -2$, $p = 1$, while keeping the spatial location of the sphere and the beam propagation direction invariant. This transformation is the mirror in the $y$ plane $\mathbf{M}_y$, which is formed combining a rotation of $\pi$ around the $y$ axis, $\mathbf{R}_{y,\pi}$, and the parity, $\Pi$. (See Appendix D for more details.)

### C. Spin/orbit torques in LG optical traps

As shown in Sec. VI B, angular momentum can be transferred to the particle in two different ways when interacting with a Laguerre–Gaussian (LG) beam: the first way is the spin angular momentum transfer, which, in this work, is referred to as "spin torque" or simply "torque" and generates the rotational motion of the particle around its own axis. The spin torque can be calculated using Eqs. (21) and (22). The second way is the orbital angular momentum transfer, which, in this work, is referred to as "orbital torque" and arises from a tangentially oriented constant force ($F_{tan}$) in the off-axis trapping configuration. More specifically, the orbital torque is calculated multiplying the value of $F_{tan}$ by the radial equilibrium position of the particle ($R_{eq}$).

In this subsection, we study the $z$ components of these two forms of angular momentum acquired by the particle. Therefore, we show the calculations of the corresponding values of $T_z$ and $F_{tan} \cdot R_{eq}$ in Fig. 4. Here, we consider a system formed by a spherical particle with a constant refractive index of $1.5 + 0.0025i$, which corresponds to an absorption coefficient of $\alpha_{abs} = 295.26$ cm$^{-1}$. In addition, this particle is trapped in water by focused LG modes with six different topological charges (from $\ell = 0$ to 5), all of them with a fixed wavelength of $\lambda = 1064$ nm and an optical power of 10 mW. Figure 4 depicts the values of $T_z$ (continuous lines) and $F_{tan} \cdot R_{eq}$ (dashed lines) when varying the optical size of the particle from $x = 2.5$ to 10. We always calculate the values at the actual equilibrium point for every particle. In some cases, this equilibrium point is on-axis or placed along the $z$ axis ($x = 0$, $y = 0$, $z \neq 0$), and, thus, $R_{eq} = 0$, while in other situations, it is on the bright ring of the Laguerre–Gaussian beam, leading to the off-axis trapping regime.

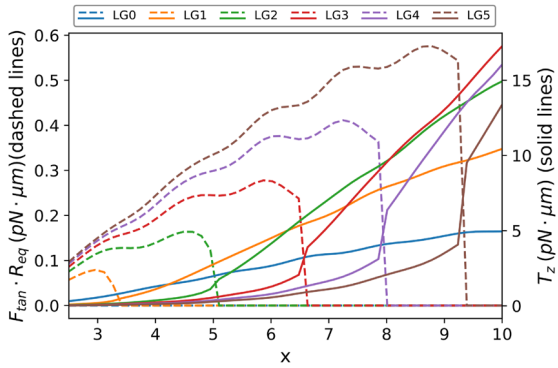

**FIG. 4.** Spin and orbital optical torques generated by focused Laguerre–Gaussian modes over an absorbing spherical particle with optical size $x$ placed at the actual trapping point. For these calculations, we considered Laguerre–Gaussian modes with topological charges from $\ell = 0$ to $\ell = 5$ focused by an objective lens with NA = 1.25. In addition, the spherical particle has a constant refractive index of $1.5 + 0.0025i$ and is surrounded by water.

A noticeable effect that can be observed in Fig. 4 is the abrupt changes in the torques when the trapping regime changes from off-axis (small particles) to on-axis trapping (large particles). This change can be easily recognized by observing the sudden drop of $F_{tan} \cdot R_{eq}$ (dashed lines) when a certain value of $x$ is reached. We can also notice that, for higher topological charge orders of the trapping LG mode, the on-axis trapping regime starts at a higher value of $x$. This effect is attributed to the increment of the radius $r_\ell$ of the ring of the trapping LG-mode at a focal plane, which increases with the topological charge as $r_\ell = \sqrt{z_\ell \ell / k}$, with $z_\ell$ being the Rayleigh range of the beam. In addition, at the switching point of the trapping regime, we observe a substantial jump in the corresponding value of $T_z$ (solid lines), modifying its scaling with respect to $x$ from quadratic to linear. These jumps correspond to an increment in the value of $T_z$ significantly higher than the one lost by the disappearing of $F_{tan} \cdot R_{eq}$ (notice the different scales for both torques). This means that the particle acquires more angular momentum when it is on-axis than when it is off-axis. In addition, notice that the transfer of orbital angular momentum to the particle is, in general, not proportional to $\ell$.

We can also calculate the maximum rotation rates of these trapping situations, using the formula for drag spin torque experienced by a sphere spinning far from any surfaces in a fluid $T_{drag}^{spin} = 8\pi\eta a^3 \omega_{rot}^{spin}$,[55] where $\eta$ is the viscosity of the fluid surrounding the particle, $a$ is the radius of the spherical particle, and $\omega_{rot}^{spin}$ is the rotation rate. The steady state of the rotation will happen when the optical torque and the drag torque are equal, leading to a maximum spin rotational velocity $\omega_{max}^{spin}$. The same logic would be applied for calculating the maximum orbital rotation rate $\omega_{max}^{orbital}$, but in this case, it would occur when the drag force of the medium is equal to the optical tangential force that generates the orbital rotation of the particle. Here, the drag orbital torque of a sphere in a fluid undergoing constant orbital rotation separated $R_{eq}$ from the beam axis would be given by Stoke's law as $T_{drag}^{orbital} = 6\pi\eta a R_{eq}^2 \, \omega_{rot}^{orbital}$.[55]

Furthermore, we can use these expressions in order to calculate the rotational frequencies of an optical trapping case that generates both rotational regimes simultaneously. For instance, the case of Fig. 4 in which the LG3 beam traps the particle with $x = 6$ in water ($\eta_{water} = 1.002 \times 10^{-3}$ N s/m$^2$) fulfills this condition. In this trapping situation, we see that the optical spin torque exerted to the particle is of 1.3 pN $\mu$m and the optical orbital torque is of 0.3 pN $\mu$m. Applying the equations above, we can estimate that the maximum spin rotation rate would be $\omega_{max}^{spin} = 48.3$ rad/s, resulting in a spin rotation frequency of $f_{max}^{spin} = 7.7$ Hz, and the maximum orbital rotation rate would be $\omega_{max}^{orbital} = 30.9$ rad/s, resulting in an orbital rotation frequency of $f_{max}^{orbital} = 4.9$ Hz.

## D. Optical trapping with LG beams near optical resonances and their relation to the scattering efficiency

We have found earlier that, in certain situations, there can appear narrow resonant-like features in the optical forces exerted over spherical particles. Here, we will further explore this situation, analyzing the longitudinal optical force $F_z$ and its relationship with the scattering efficiency $Q_{sca}$, which is calculated as

$$Q_{sca} = \sum_{j,m_z,p} |g_{j,m_z,p}|^2 \left( |\alpha_j|^2 + |\beta_j|^2 \right). \tag{24}$$

The key element for this discussion is the realization that centered Laguerre–Gaussian (LG) beams do not excite modes with $j < |m_z^*|$. For LG beams with $m_z^* = \ell + p > 1$, this fact effectively clears dipolar modes in the optical response of the object. This is important as the dipolar contribution typically contributes across the entire spectrum. Higher multipolar modes, on the other hand, usually exhibit sharp resonances that decay rapidly outside these resonances. The combination of these two effects simplifies the longitudinal force shown in Eq. (18), as there is no interference between the $B_{j-1}$ and $B_{j+1}$ terms when the $B_j$ mode resonates. In other words, the longitudinal optical force $F_z$ simplifies significantly due to the absence of interference between adjacent electromagnetic modes.

In Fig. 5, we show the longitudinal forces exerted by focused Laguerre–Gaussian modes on spherical particles with refractive index $n = 3.5$ in air. As expected, strong resonances arise when using modes with $\ell > 0$, resulting in a finer agreement between $F_z$ and $Q_{sca}$ for LG beams with topological charge $\ell > 1$. Interestingly, in these situations, the scattering efficiency $Q_{sca}$ is also proportional to the longitudinal force $F_z$, given that both of them are proportional to $|B_j|^2$ near the Mie resonances. For $\ell = 0$ (Gaussian beam), however, the dipolar contribution persists across the spectrum, breaking the correspondence between $F_z$ and $Q_{sca}$.

These distinctive features can be used to measure the spectral positions of resonant peaks by monitoring the particle's position along the $z$-axis. As the wavelength of the incident beam is tuned, a particle's sudden displacement in the positive $z$-direction at resonant optical sizes could provide a simple and effective method for observing higher-order Mie resonances.

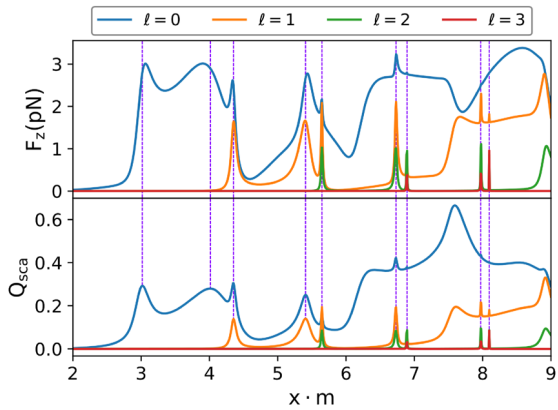

**FIG. 5.** Longitudinal forces ($F_z$) and scattering efficiency ($Q_{sca}$) generated by focused Laguerre–Gaussian beams with helicity $p = 1$ and topological charge order $\ell$ over a spherical particle with n = 3.5. The values of the optical force have been calculated as a function of the optical size of the particle multiplied by its refractive index contrast $x \cdot m$, ranging from 2 to 9. Here, it is possible to observe the suppression of the contribution of the lowest multipolar orders and the sharpening of the resonances when the topological charge of the incident beam increases.

### E. The influence of helicity in longitudinal optical forces

While, in Secs. VI A–VI D, we explored the role of the total angular momentum for higher order Laguerre–Gaussian (LG) modes, in this subsection, we want to further understand the relationship of helicity to optical forces, if there is any. However, this situation is more contrived and difficult to study experimentally because the helicity is better controlled in the paraxial regime. In the paraxial regime, changing the circular polarization of the incident field has two effects on the multipolar decomposition at the focal point, as seen in Sec. III A: (1) the multipolar modes have a different $p$, and (2) they also have a different $m_z^*$. For example, if we simply change the circular polarization of a typical paraxial incident Gaussian beam from, say, $p = -1$ to $p = 1$, this would not change the longitudinal optical force at all, because the beam with flipped helicity would also have a flipped $z$ component of total angular momentum in $m_z^*$. To address this issue, we can consider using LG beams. For instance, we can construct a LG beam with $p = 1$ and $m_z^* = -1$, implying that it has the same total angular momentum in $z$ as the initial Gaussian beam with $p = -1$, while its helicity has been flipped as intended. However, this beam would have completely different multipolar coefficients, thereby inducing more than just a pure helicity modification.

Therefore, if we want to understand the unique role of flipping the helicity of the multipolar decomposition without affecting any other parameter, we need to resort to a more "artificial" scenario. We will start with a focused Gaussian beam with a given helicity, with the particle placed on-focus, and then we keep the multipolar content and the $m_z^*$ of the incident beam constant, but flip its helicity. From the perspective of the multipolar decomposition used in this work [see Eq. (8)], we consider the difference in forces from these two electromagnetic fields,

$$\mathbf{E}_{in}^{on} = E_0 \sum_{j=1}^{\infty} g_j^{on} \mathbf{A}_{j,1}^{+1}, \tag{25}$$

$$\mathbf{E}_{in}^{on} = E_0 \sum_{j=1}^{\infty} g_j^{on} \mathbf{A}_{j,1}^{-1}. \tag{26}$$

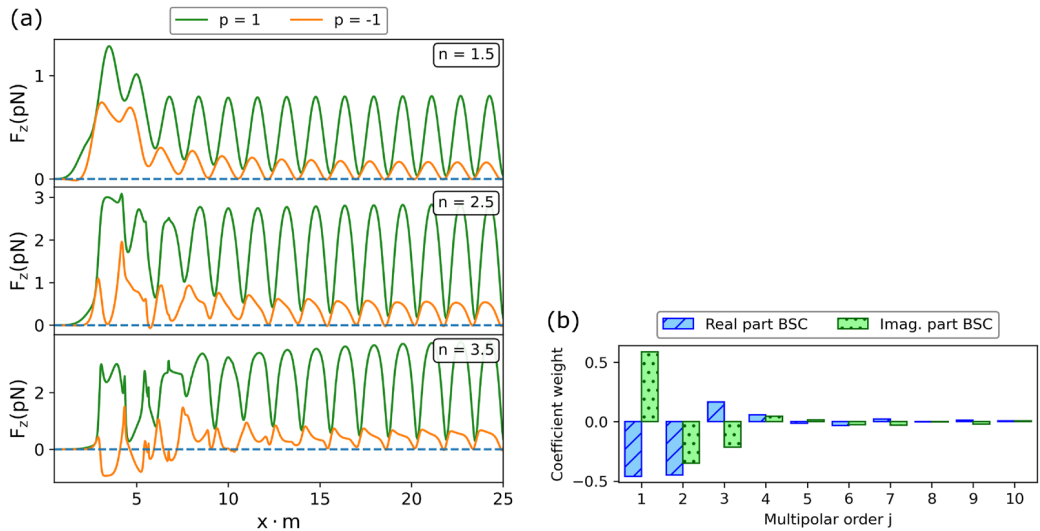

**FIG. 6.** (a) Longitudinal optical forces ($F_z$) exerted over spherical particles with three different refractive indices. The values of the optical force have been calculated as a function of the optical size of the particle multiplied by its refractive index contrast $x \cdot m$, ranging from 0.5 to 25. These calculation consider two focused beams, one with the multipolar description of (25) (green line) and the other with the one of (26) (orange line). Both contain the same multipolar coefficients shown in panel (b) and have an optical power of 1 mW. (b) Multipolar coefficients of the left polarized Gaussian beam. This set of coefficients are calculated using the solution of Eq. (9), considering a focusing lens with NA = 0.95 and a filling factor of 0.96, focusing a beam with $\lambda = 976$ nm and topological charge $\ell = 0$ in air.

The first scenario analyzed in Fig. 6(a) (green line) shows the longitudinal optical forces produced by a focused Gaussian beam with $p = 1$ described by Eq. (25), with the multipolar weights shown in Fig. 6(b). In the second scenario (orange line), we observe the longitudinal forces for a beam with the same multipolar coefficients and $m_z^*$ but with helicity $p = -1$, described by Eq. (26). Note that, in this second case, we are not considering a conventional focused beam, and it could be quite challenging to generate it experimentally. In particular, if we integrate the Poynting vector of this beam, we find that it is composed of two counter-propagating components, with 22% of the total beam energy flowing in the negative direction of the $z$ axis. The optical force values shown in Fig. 6 are calculated considering that these two beams impinge on the center of a dielectric spherical particle (on-focus configuration) with refractive indices $n = 1.5, 2.5$, and $3.5$ and surrounded by air. In addition, we consider that the optical power of the beam is 1 mW and that it is focused with a lens of NA = 0.95.

In Fig. 6, one can observe that the longitudinal optical forces generated by these two beams are clearly different. This is not surprising, considering that the beam with $p = -1$ does not resemble a normal focused Gaussian beam with that helicity. Another remarkable feature is the oscillating behavior of the longitudinal optical forces along $x \cdot m$. This is attributed to the varying coupling efficiency of the multipolar modes of the incoming light beam with the spherical particle as its optical size changes, which has been observed in other studies.[56] Note also that the period of the oscillations remains constant for the three different refractive index values considered in these calculations. In addition, as n increases, we can identify the generation of extra ripples associated with Mie resonances for the lowest values of $x \cdot m$. Furthermore, in the case of $p = 1$ and $n = 3.5$, from $x \cdot m \simeq 1$ to 5, we can distinguish a structure similar to the scattering cross section of a dipolar Si sphere reported in Ref. 57, which has been discussed in Sec. VI D. In the same spectral region for the case of $n = 3.5$, we observe that the beam with $p = -1$ is able to generate negative forces, which could be explained by the fact that 22% of the total beam power is propagating in the negative direction of the $z$ axis.

## VII. CONCLUSIONS

In this study, we have derived and analyzed analytical expressions for the optical forces and torques exerted on spherical particles illuminated by Laguerre–Gaussian (LG) beams, using the well-defined helicity multipolar decomposition. We have introduced the multifunctional program MOFT, available on GitHub, which enables us to generate optical trapping stability maps that predict the stability of LG optical traps across a wide range of beam and particle parameters.

By expanding the electromagnetic fields into well-defined helicity multipoles, our theoretical framework provides an efficient solution to the BSC for describing focused LG beams that satisfy Maxwell's equations. In addition, the representation of displaced beams used in this work enables a comprehensive analysis of the optical forces and torques in LG optical traps with spherical particles, while the scatterer coefficients remain invariant.

In addition, since most circularly polarized beams focused by a microscope objective can be approximated as having a well-defined helicity value, this quantity becomes a powerful characterization for such beams, further reinforcing the suitability of our framework. More specifically, the well-defined helicity multipolar basis allows us to identify how the cylindrical symmetry of the system, together with fundamental beam parameters such as helicity $p$ and topological charge $\ell$, affect the multipolar content of the beam. This insight enables us to distinguish and characterize different optical trapping regimes, such as on- or off-focus and on- or off-axis configurations.

Furthermore, our theoretical framework has allowed us to systematically analyze the impact of sign modifications in helicity $p$ and topological charge $\ell$ on the resulting optical forces and torques, revealing the intricate interplay between these quantities and the $z$ component of total angular momentum $m_z$ in LG beams. This interplay shapes the dynamics of the trapped particle, demonstrating peculiar behaviors such as the exceptional nature of tangential torque and its dependence on $\ell$ and $p$.

We observed notable differences in longitudinal optical forces arising from pure helicity modifications, which emphasize the importance of helicity in determining the optical forces. Moreover, our study has shown that LG beams can isolate Mie resonances, providing a novel method for identifying the spectral positions of higher-order resonant peaks through the measurement of longitudinal forces.

Overall, this work not only advances our theoretical understanding of optical forces and torques in the context of structured light fields such as LG modes but also highlights the effectiveness of the well-defined helicity multipolar basis in describing tightly focused beams and their interaction with trapped particles. In this regard, specific experimental studies in which the helicity of the incident beam plays a crucial role, such as the filtering of harmful enantiomers in biomedical applications[58–65] and the control and manipulation of light-driven nanorotors,[66–68] can greatly benefit from the theoretical tools provided in this work. In any case, the theoretical insights provided in this work have the potential to open new avenues for experimental exploration and technological development in optical trapping and manipulation.

## ACKNOWLEDGMENTS

Iker Gómez-Viloria, Gabriel Molina-Terriza, Enrique Ayllón García, Quimey Pears Stefano, Jon Lasa-Alonso, and Martín Molezuelas–Ferreras acknowledge the support from CSIC Research Platform on Quantum Technologies PTI-001, from IKUR Strategy under the collaboration agreement between Ikerbasque Foundation and DIPC/MPC on behalf of the Department of Education of the Basque Government and from Project Nos. EQC2018-004060-P and PID2022-143268NB-I00 of Ministerio de Ciencia, Innovación y Universidades.

Jorge Olmos-Trigo acknowledges the support from the Juan de la Cierva fellowship under Grant No. FJC2021-047090-I of MCIN/AEI/10.13039/501100011033 and NextGenerationEU/PRTR and from the Spanish ministry under Grant Nos. PID2022-137569NB-C43 and PID2022-143268NBI00.

Martín Molezuelas-Ferreras acknowledges the financial support from the Spanish Ministerio de Ciencia, Innovación y Universidades, through the FPI PhD Fellowship under Grant No. FIS-2017-87363-P.

## AUTHOR DECLARATIONS

### Conflict of Interest

The authors have no conflicts to disclose.

### Author Contributions

**Iker Gómez-Viloria**: Conceptualization (equal); Data curation (equal); Formal analysis (equal); Investigation (equal); Methodology (equal); Software (equal); Validation (equal); Visualization (equal); Writing – original draft (equal); Writing – review & editing (equal). **Enrique Ayllón García**: Data curation (supporting); Formal analysis (supporting); Visualization (supporting); Writing – original draft (supporting); Writing – review & editing (supporting). **Jorge Olmos-Trigo**: Conceptualization (supporting); Formal analysis (supporting); Supervision (supporting); Writing – original draft (supporting); Writing – review & editing (supporting). **Quimey Pears Stefano**: Data curation (supporting); Software (supporting); Supervision (supporting); Validation (supporting). **Jon Lasa-Alonso**: Conceptualization (supporting); Formal analysis (supporting); Supervision (equal); Writing – original draft (supporting); Writing – review & editing (supporting). **Martín Molezuelas-Ferreras**: Conceptualization (supporting); Formal analysis (supporting); Writing – original draft (supporting); Writing – review & editing (supporting). **Gabriel Molina-Terriza**: Conceptualization (equal); Formal analysis (equal); Funding acquisition (equal); Investigation (equal); Project administration (equal); Resources (equal); Supervision (equal); Writing – original draft (equal); Writing – review & editing (equal).

### DATA AVAILABILITY

The data that support the findings of this study are available from the corresponding author upon reasonable request.

## APPENDIX A: ELECTROMAGNETIC FIELDS IN THE ELECTRIC AND MAGNETIC MULTIPOLAR BASIS

We will first proceed by developing the scattering theory using standard electric ($\mathbf{A}_{j,m_z}^{(e)(n)}(\mathbf{r})$) and magnetic ($\mathbf{A}_{j,m_z}^{(m)(n)}(\mathbf{r})$) multipoles to describe the incident and scattered electromagnetic fields. We will follow the definitions given by Rose in Eqs. (2.63) and (2.64) of Ref. 32,

$$\mathbf{A}_{j,m_z}^{(m)(n)}(\mathbf{r}) = -z_j^{(n)}(kr)\mathbf{T}_{jj}^{m_z}(\hat{\mathbf{r}}) \tag{A1}$$

and

$$\mathbf{A}_{j,m_z}^{(e)(n)}(\mathbf{r}) = -\sqrt{\frac{j}{2j+1}}z_{j+1}^{(n)}(kr)\mathbf{T}_{jj+1}^{m_z}(\hat{\mathbf{r}})$$
$$+ \sqrt{\frac{j+1}{2j+1}}z_{j-1}^{(n)}(kr)\mathbf{T}_{jj-1}^{m_z}(\hat{\mathbf{r}}), \tag{A2}$$

where

$$\mathbf{T}_{jl}^{m_z}(\hat{\mathbf{r}}) = \sum_{\mu=-1}^{\mu+1}\langle 1,\mu\,;l,m_z-\mu\mid j,m_z\rangle\, Y_l^{m_z+\mu}(\hat{\mathbf{r}})\,\boldsymbol{\xi}_{-\mu} \tag{A3}$$

are the vector spherical harmonics.

Here, we can find the Clebsch–Gordan coefficients with the notation $\langle j_1,m_1\,;\,j_2,m_2\mid J,M\rangle$ and the scalar spherical harmonics $Y_j^{(m/e)(n)}(\hat{\mathbf{r}})$.[39] The multipole modes $\mathbf{A}_{j,m_z}^{(m/e)(n)}$ are characterized by being eigenmodes of the total angular momentum $\mathbf{J}^2$ and of the z component of the total angular momentum $J_z$, with $j(j+1)$ and $m_z$ being their respective eigenvalues. Typically, $j$ is also referred to as the multipolar order. The multipoles depend on the position vector $\mathbf{r}$, which, in Cartesian coordinates, reads $\mathbf{r} = (x, y, z)$. The unitary vectors $\boldsymbol{\xi}_\mu$ determine the polarization of the field with the relations $\boldsymbol{\xi}_{+1} = -(\hat{\mathbf{x}} + i\hat{\mathbf{y}})/\sqrt{2}$, $\boldsymbol{\xi}_{-1} = (\hat{\mathbf{x}} - i\hat{\mathbf{y}})/\sqrt{2}$, and $\boldsymbol{\xi}_0 = \hat{\mathbf{z}}$. Moreover, $z_j^{(n)}(kr)$ denotes spherical Bessel functions, with $n$ being the kind of spherical Bessel functions, which will be determined by the type of electromagnetic field we are dealing with. That is, $z_j^{(1)}(kr) = j_j(kr)$ for the incident electromagnetic field, while $z_j^{(3)}(kr) = h_j^{(2)}(kr)$ for the scattered electromagnetic field, with $j_j(kr)$ and $h_j^{(2)}(kr)$ being the spherical Bessel and the spherical outgoing Hankel functions, respectively (this notation will often be omitted in the rest of this appendix). Please also note that $k = \sqrt{\mu\varepsilon}k_0$ is the wavenumber of the electromagnetic field in the surrounding medium, with $\varepsilon$ and $\mu$ being its electric permittivity and magnetic permeability, respectively, and $k_0$ being the wavenumber in vacuum.

Magnetic and electric multipolar modes differ in their behavior under parity transformations: $\Pi\mathbf{A}_{j,m_z}^{(e)} = (-1)^j\mathbf{A}_{j,m_z}^{(e)}$ and $\Pi\mathbf{A}_{j,m_z}^{(m)} = (-1)^{j+1}\mathbf{A}_{j,m_z}^{(m)}$, where $\Pi$ is the parity operator.[11] They are related by the following transformation:

$$\mathbf{A}_{j,m_z}^{(e)(n)}(\mathbf{r}) = -i\Lambda\mathbf{A}_{j,m_z}^{(m)(n)}(\mathbf{r}), \tag{A4}$$

where $\Lambda = (1/k)\nabla\times$ is the helicity operator for monochromatic fields.[11] Note that $\Lambda^2 = I$, with $I$ being the $3 \times 3$ identity matrix.

Any incident electromagnetic field can be described as a superposition of electric and magnetic multipoles of type $x = 1$, i.e., $\mathbf{A}_{j,m_z}^{(m/e)(1)}$. The amplitude of the multipolar modes is typically called Beam Shape Coefficient (BSC) and is denoted by $g_{j,m_z}^{(e)}$ and $g_{j,m_z}^{(m)}$[28]

$$\mathbf{E}_{in} = E_0\sum_{j=1}^{\infty}\sum_{m_z=-j}^{j}g_{j,m_z}^{(m)}\mathbf{A}_{j,m_z}^{(m)(1)} - g_{j,m_z}^{(e)}\mathbf{A}_{j,m_z}^{(e)(1)}, \tag{A5}$$

where $E_0$ is the normalized amplitude of the electric incident field for this decomposition.

The scattering by a spherical object is analytically described by Mie theory and can be written in terms of electric and magnetic Mie coefficients,[69,70] which for non-magnetic spheres ($\mu = 1$) read

$$a_j = \frac{\mu_m m\,\psi_j'(x)\,\psi_j(mx) - \mu_p\,\psi_j(x)\,\psi_j'(mx)}{\mu_m m\,\xi_j'(x)\,\psi_j(mx) - \mu_p\,\xi_j(x)\,\psi_j'(mx)},$$
$$b_j = \frac{\mu_p m\,\psi_j(x)\,\psi_j'(mx) - \mu_m\,\psi_j'(x)\,\psi_j(mx)}{\mu_p m\,\xi_j(x)\,\psi_j'(mx) - \mu_m\,\xi_j'(x)\,\psi_j(mx)}, \tag{A6}$$

where $x_s = ka$ is the optical size of the sphere, with $a$ being its radius, and $m = n/n_m$ is the refractive index contrast of the sphere's material, with $n$ and $n_m$ being the refractive index of the sphere and the surrounding medium, respectively. Meanwhile, we have that $\psi_j(x) = x\,j_j(x)$ and $\xi_j(x) = x\,h_j(x)$ and that $\mu_p$ and $\mu_m$ are the magnetic permeability inside and outside the particle, respectively.

Therefore, the incident field in Eq. (A5) impinging on a sphere produces the scattered electric field,

$$\mathbf{E}_{sc} = -E_0 \sum_{j=1}^{\infty} \sum_{m_z=-j}^{j} g_{j,m_z}^{(m)} b_j \mathbf{A}_{j,m_z}^{(m)(3)} - g_{j,m_z}^{(e)} a_j \mathbf{A}_{j,m_z}^{(e)(3)}, \qquad (A7)$$

and the total electric field is the superposition of the incident and scattered fields: $\mathbf{E}_{tot} = \mathbf{E}_{in} + \mathbf{E}_{sc}$.

The incident and scattered magnetic fields can be derived from the relation $Z\mathbf{H} = -i\Lambda\mathbf{E}$, where $Z = \sqrt{\mu_0/\varepsilon_0}$ is the impedance of the electromagnetic wave, with $\varepsilon_0$ being the permittivity of vacuum and $\mu_0$ being the vacuum permeability, which is readily done using Eq. (A4).

## APPENDIX B: LAGUERRE–GAUSSIAN MODES

The general expression for paraxial LG modes is given by

$$LG_{l,q}(\rho) = N_{l,q}^{LG} \exp\left[\frac{-\rho^2}{w(z)^2}\right] \rho^{|l|} L_q^{|l|}\left(\frac{2\rho^2}{w(z)^2}\right) \left(\frac{\sqrt{2}}{w(z)}\right)^{|l|+1}$$
$$\times \exp\left[i\left(kz\left(1 - \frac{\rho^2}{2(z^2 + z_0^2)}\right)\right.\right.$$
$$\left.\left. + (2q + l + 1)\,\tan^{-1}(z/z_0)\right)\right], \qquad (B1)$$

where $L_q^{|l|}$ are the generalized Laguerre polynomials, with $l$ being the azimuthal and $q$ being the radial mode numbers. The $N_{l,q}^{LG}$ are the normalization function of the Laguerre–Gaussian mode, which would be

$$N_{l,q}^{LG} = \sqrt{\frac{q!}{\pi(q + |l|)!}}. \qquad (B2)$$

## APPENDIX C: OPTICAL FORCES EXERTED BY BEAMS WITH WELL-DEFINED HELICITY OVER SPHERICAL PARTICLES

The expressions for optical forces in Sec. IV are generalized to any helicity state by summing multipoles of both components, now $p = +1$ and $p = -1$, in the incident field. If a single helicity state is considered, a simplified version of these expressions can be obtained, describing the optical forces produced by a circularly polarized beam that has been focused.

In addition, if we assume a spherical scatterer, we can use the Mie coefficients or, equivalently, the coefficients of Eq. (6). This would reduce these calculations to terms involving the BSC $A_{j,m_z,p}$ = $g_{j,m_z,p}$ according to Eq. (17) and a few parameters that define the spherical particle. Thus, the expressions of the different components of the forces under these two considerations are

$$F_z = \frac{\varepsilon}{2\,k^2} \sum_{j=1}^{\infty} \sum_{m_z=-j}^{j} \frac{1}{j+1}\left[\sqrt{w_j}\,w_{j,m_z}^{(1)}\,\mathrm{Im}\left(A_{j,m_z,p} A_{j+1,m_z,p}^* \,\mathrm{T}_j^{(1)}\right)\right.$$
$$\left. - \frac{p\,m_z}{j}\,|A_{j,m_z,p}|^2\,\mathrm{T}_j^{(2)}\right] \qquad (C1)$$

and

$$\begin{bmatrix} F_x \\ F_y \end{bmatrix} = \begin{bmatrix} \mathrm{Re} \\ \mathrm{Im} \end{bmatrix} \frac{i\,\varepsilon}{4\,k^2} \sum_{j=1}^{\infty} \sum_{m_z=-j}^{j} \frac{1}{j+1}$$
$$\times \left( \sqrt{w_j}\,w_{j,m_z}^{(2)}\,A_{j+1,m_z,p} A_{j,m_z+1,p}^* \,\mathrm{T}_j^{(1)*}\right.$$
$$+ \sqrt{w_j}\,w_{j,m_z}^{(3)}\,A_{j,m_z,p} A_{j+1,m_z+1,p}^* \,\mathrm{T}_j^{(1)}$$
$$\left. + p\,\frac{2i\,w_{j,m_z}^{(4)}}{j}\,A_{j,m_z,p} A_{j,m_z+1,p}^* \,\mathrm{T}_j^{(2)}\right), \qquad (C2)$$

where

$$\mathrm{T}_j^{(1)} = \alpha_j + \alpha_{j+1}^* + 2\left(\alpha_j\alpha_{j+1}^* + \beta_j\beta_{j+1}^*\right) \qquad (C3)$$

and

$$\mathrm{T}_j^{(2)} = |\alpha_j|^2 - |\beta_j|^2 + \mathrm{Re}(\alpha_j) \qquad (C4)$$

wrap all the parameters of the spherical particle.

## APPENDIX D: SYMMETRY TRANSFORMATIONS AND THEIR RELATIONS TO FORCES AND TORQUES

As mentioned in Sec. VI B, the mirror transformation in the $y$ plane $\mathbf{M}_y$ is a combination of a rotation of $\pi$ around the $y$ axis, $\mathbf{R}_y(\pi)$, and the parity, $\Pi$. Let us see how these transformations affect the values of the optical forces and torques in the optical trapping system of Sec. VI B.

First, the optical forces transferred to the particle emerge as a consequence of variations in the linear momentum carried by the incident beam, so let us apply the $\mathbf{M}_y$ transformation to the linear momentum operator $\mathbf{P}$ in order to infer how the optical forces are affected by this transformation,

$$\mathbf{M}_y \cdot \mathbf{P} = \mathbf{R}_y(\pi) \cdot \Pi \cdot [P_x, P_y, P_z]$$
$$= \mathbf{R}_y(\pi) \cdot [-P_x, -P_y, -P_z]$$
$$= [P_x, -P_y, \ P_z]. \qquad (D1)$$

A similar reasoning can be employed for the optical torques, now applying the $\mathbf{M}_y$ transformation to the angular momentum operator $\mathbf{L} = \mathbf{r} \times \mathbf{P}$, so that we have

$$\mathbf{M}_y \cdot \mathbf{L} = \mathbf{R}_y(\pi) \cdot \Pi \cdot [(r_yP_z - r_zP_y)\hat{x}, (r_zP_x - r_xP_z)\hat{y}, (r_xP_y - r_yP_x)\hat{z}]$$
$$= \mathbf{R}_y(\pi) \cdot [(r_yP_z - r_zP_y)\hat{x}, (r_zP_x - r_xP_z)\hat{y}, (r_xP_y - r_yP_x)\hat{z}]$$
$$= [-(r_yP_z - r_zP_y)\hat{x}, (r_zP_x - r_xP_z)\hat{y}, -(r_xP_y - r_yP_x)\hat{z}]$$
$$= [-L_x, \ L_y, -L_z]. \qquad (D2)$$

The results obtained in Eqs. (D1) and (D2) are in complete agreement with the results of Fig. 3, ratifying the way each component of the optical forces and torques behaves when a simultaneous flip of the sign of $\ell$ and $p$ is performed. First, as shown in Eq. (D1), this transformation changes the sign of the $y$ component of the optical forces, and second, as Eq. (D2) confirms, this transformation also changes the sign of the optical torque components $x$ and $z$, while the sign of $y$ remains invariant.

28 January 2026 19:28:14